\documentclass{SCGEa}

\begin{document}

\def\bm{\boldsymbol}

\def\dl{\displaystyle}
\def\tlj{\end{document}}
\def\d{{\rm d}}
\def\e{{\rm e}}
\def\i{{\rm i}}

% The author doesn't need fill in it.
\Year{} %
\Month{} %
\Vol{} %  卷号
\No{} %  期号
\BeginPage{1} % 起页码
\AuthorMark{{\rm QIAN Y. Z.}}  %(11月注释：页眉上的作者)
\AuthorMarkCite{{\rm Qian Y Z.}~} %(11月注释：citation中的作者)
\DOI{} % The author doesn't need fill in it.
\ArtNo{x}

% \title[short text for running head]{full title}{comments for title}
\title[Neutrinos, supernovae, and the origin of the heavy elements]
{Neutrinos, supernovae, and the origin of the heavy elements}%标题 {有黑体的时候需要将标题复制在中括号里面，使得引用条显示白体。没有黑体的时候中括号可以删掉}

\author[*1,2]{QIAN YONG-ZHONG}{}
%\footnote{*Corresponding author (email: qian@physics.umn.edu)}%手动E-mail地址

\address[1]{School of Physics and Astronomy, University of Minnesota, Minneapolis, Minnesota 55455, USA}
\address[2]{Tsung-Dao Lee Institute, Shanghai 200240}

\maketitle \vspace{-3.5mm}{\footnotesize\begin{center} Received ; accepted ; published online %收稿日期
\end{center}}\vspace*{-5mm}

%     Abstract is required.
\begin{center}
\rule{16.5cm}{0.4pt}
\parbox{16.5cm}
{\begin{abstract} Stars of $\sim 8$--$100M_\odot$ end their lives as core-collapse supernovae (SNe). 
In the process they emit a powerful burst of neutrinos, produce a variety of elements, and leave behind 
either a neutron star or a black hole. The wide mass range for SN progenitors results in diverse 
neutrino signals, explosion energies, and nucleosynthesis products. A major mechanism to produce 
nuclei heavier than iron is rapid neutron capture, or the $r$ process. This process may be connected 
to SNe in several ways. A brief review is presented on current understanding of neutrino emission, 
explosion, and nucleosynthesis of SNe.
\end{abstract}}
\end{center}\vspace*{-0.6cm}

\begin{center}
\parbox{16.5cm}
{\bf\jiuhao neutrino; supernova; nucleosynthesis; the $r$ process}%关键词
\end{center}

\begin{center}
{}%分类号
\CITA    %%(11月注释：Citation内容自动生成)
%\Cit{~~~???, et al. ???. Sci China-Phys Mech Astron, 2014, 57: 1--6, doi:}%%(11月注释：Citation内容需手动填写)
\end{center}

\textwidth=178truemm \textheight=236truemm%%%%%%新版式要加上

%%%%%%%%%%%%%%%%%%%%%%%%%%%%%%%%%%%%%%%%%%%%%%%%%%%%%%%%%%%%
\wuhao\vspace*{1.5mm}

\begin{multicols}{2}

%%%%%%%%%%%%%%%%%%%%%%%%%%%%%%%%%%%%%%%%%%%%%%%%%%%%%%%%%%%%
%% Text of article.
%%%%%%%%%%%%%%%%%%%%%%%%%%%%%%%%%%%%%%%%%%%%%%%%%%%%%%%%%%%%
%    Section headings
\renewcommand{\baselinestretch}{1.08} \baselineskip 12.2pt\parindent=10.8pt

\renewcommand{\thefootnote}

\section{Introduction}
After baryogenesis in the early universe and when the temperature drops 
to $T\sim 100$~MeV, the only baryons present are neutrons and protons. 
Because matter at such a high temperature is in thermal equilibrium, 
the neutron-to-proton ratio $n/p$ is determined by their mass difference 
$\Delta=m_n-m_p=1.293$~MeV through the Boltzmann factor:
\begin{equation}
n/p = \exp(-\Delta/T).
\end{equation}
It is a remarkable achievement of the standard model of particle physics 
that this mass difference can now be calculated from first principles to 
within 20\% \cite{npmass}.
Because a neutron is heavier than a proton, the equilibrium abundance 
shifts more and more towards protons as $T$ drops. This shift is 
accomplished by the competition among the weak interactions 
interconverting neutrons and protons:
\begin{align}
\nu_e + n&\rightleftharpoons p + e^-,\label{eq-np}\\
\bar\nu_e + p&\rightleftharpoons n + e^+.\label{eq-pn}
\end{align}
Because the rates of these interactions also decrease with $T$, 
the neutron-to-proton ratio eventually freezes out at $T\sim 1$~MeV. 
As the universe cools further, big bang nucleosynthesis fuses 
neutrons and protons into light nuclei such as $^2$H, $^3$H, $^3$He, 
$^4$He, $^7$Li, and $^7$Be.

Under the influence of gravity, big bang debris containing mostly 
protons condenses into stars, which shine 
by burning protons into heavier nuclei and provide the newly-synthesized 
products to the interstellar medium when they die. Therefore, the 
next generation of stars formed from this medium are enriched 
beyond the big bang composition. This cycle repeats as generation 
after generation of stars are born, lead luminous lives, die glorious 
deaths, and in the process convert primordial baryons into nuclei of 
the entire periodic table. To quantify this picture of cosmic alchemy, 
we can compare the composition of big bang debris with that in the 
sun, which formed $\approx 9$~Gyr after the big bang. The dominant 
products of big bang nucleosynthesis are protons and $^4$He with mass 
fractions of $\approx 75\%$ and $\approx 25\%$, respectively. The mass 
fractions of protons, $^4$He, and nuclei for the rest of the elements in the 
sun are $\approx 71.1\%$, $\approx 27.4\%$, and $\approx 1.5\%$, 
respectively. Clearly, the net effect of stellar processing is to convert 
protons into nuclei containing both protons and neutrons. Although 
the actual processes involve many steps, we can state in general 
that this end result must be achieved with the help of weak interactions 
of the following types:
\begin{align}
(Z,N)&\to(Z-1,N+1)+e^++\nu_e,\label{eq-b+}\\
e^-+(Z,N)&\to(Z-1,N+1)+\nu_e,\label{eq-ec}\\
\bar\nu_e+(Z,N)&\to(Z-1,N+1)+e^+,\label{eq-nucap}
\end{align}
where $(Z,N)$ indicates a nucleus with $Z$ protons and $N$ neutrons.

The following examples serve to illustrate the critical roles of the above weak 
interactions in providing neutrons for making nuclei. 
The $\beta^+$ decay of $^{13}$N is of the type in Eq.~(\ref{eq-b+})
and is the crucial step in the reaction sequence
$^{12}{\rm C}(p,\gamma)^{13}{\rm N}(e^+\nu_e)^{13}{\rm C}(\alpha,n)^{16}{\rm O}$
that provides a major neutron source for $s$-process nucleosynthesis
in stars of $\sim 1$--$3\,M_\odot$ (see e.g., Ref.~\cite{sproc} for a review).
The reverse reaction in Eq.~(\ref{eq-np}) is of the type in Eq.~(\ref{eq-ec}) and
is responsible for converting the Fe core of a massive ($\gtrsim 8\,M_\odot$) star 
into a neutron star (NS).
If the NS is formed in a binary with another NS or a black hole (BH) as its companion,
then it can be tapped as a powerful neutron source for $r$-process
nucleosynthesis through its disruption during the 
merger with its companion (see Sec.~\ref{sec-nsm}). The forward reaction in
Eq.~(\ref{eq-pn}) is of the type in Eq.~(\ref{eq-nucap}) and plays an important
role in supernova nucleosynthesis (see Sec.~\ref{sec-wind}). 
The above discussion shows that neutrinos are intimately associated 
with the origin of the elements. 

%\noindent\rule{2.5cm}{0.4pt}\\[0.1mm]{\qihao *Corresponding author (email: qian@physics.umn.edu)\vspace{-1mm}}%手动E-mail地址

\section{The $r$ process and supernovae}
\label{sec-rproc}
Before further discussing the roles of neutrinos in nucleosynthesis, 
we first describe how heavy nuclei are made by capturing neutrons. 
There are two prominent sets of peaks in the abundance distribution of 
nuclei heavier than Fe in the solar system. One set contains nuclei such 
as $^{138}$Ba and $^{208}$Pb with magic neutron numbers 82 and 126, 
respectively. These are produced by the so-called slow neutron-capture 
($s$) process. Once stable nuclei with magic neutron numbers are 
produced by the $s$ process, they are hard to destroy due to their small 
neutron-capture cross sections. So they pile up and form peaks.
In contrast to the $s$ process whose path stays close to stable nuclei,
the so-called rapid neutron-capture ($r$) process initially produces nuclei far from 
stability. This is because the neutron density in the $r$-process environment
is so high that neutron capture on the unstable nuclei produced occurs much 
faster than their $\beta$ decay. Unstable nuclei with magic neutron numbers
also form peaks because they are relatively more stable.
On exhaustion of neutrons, these nuclei $\beta$ decay to stability and give 
rise to the peaks at mass numbers $A\sim 130$ and 195, respectively, in 
the solar abundance distribution. 

The $r$ process (see e.g., Refs.~\cite{qian2003,qian2007,arnould,qian2014} for reviews) 
has a lot to do with the death of a massive star in 
a core-collapse supernova (SN). The connection between such an SN and 
the formation of an NS was proposed by Baade and Zwicky \cite{baade1}
shortly after the discovery of the neutron by Chadwick in 1932. They observed 
an extremely bright SN and found that the net energy of radiation
was enormous \cite{baade2}. They also found that a comparable amount of energy
from each past SN could power the cosmic rays \cite{baade1}.
To account for the required energy in each case, they made the following proposal
\cite{baade1}.
``With all reserve we advance the view that a super-nova represents the transition of 
an ordinary star into a {\it neutron star}, consisting mainly of neutrons. Such a star may 
possess a very small radius and an extremely high density. As neutrons can be packed 
much more closely than ordinary nuclei and electrons, the `gravitational packing' energy 
in a {\it cold} neutron star may become very large, and, under certain circumstances, may 
far exceed the ordinary nuclear packing fractions.''

We now know that the total amount of energy emitted in photons 
by a typical SN is $\sim 10^{49}$~ergs and the kinetic energy of the SN debris is
$\sim 10^{51}$~ergs. By comparison, the gravitational binding energy $E_G$ 
of an NS is
\begin{equation}
E_G\sim\frac{3}{5}\frac{GM_{\rm NS}^2}{R_{\rm NS}}\sim
3\times 10^{53}\left(\frac{M_{\rm NS}}{1.4M_\odot}\right)^2
\left(\frac{10\ {\rm km}}{R_{\rm NS}}\right)\ {\rm ergs},
\label{eq-eg}
\end{equation}  
where nominal values of the NS mass $M_{\rm NS}$ and radius $R_{\rm NS}$ 
are indicated. Although Baade and Zwicky did not give an explicit estimate of
$E_G$, which they referred to as the ``gravitational packing'' energy, they correctly
suggested that this energy may far exceed the binding energy released in nuclear 
reactions, which they meant by ``ordinary nuclear packing fractions.'' Indeed, 
$E_G$ corresponds to $\sim 100$~MeV/nucleon, which is much higher than the 
typical nuclear binding energy of $\sim 8$~MeV/nucleon.
Furthermore, the insightful association of NS formation and cosmic-ray 
production with SNe by Baade and Zwicky has been put on much firmer grounds.

\subsection{SN explosion}
Only massive stars of $\gtrsim 8M_\odot$ can become SNe. A star of 
$\sim 8$--$9M_\odot$ develops an O-Ne-Mg core at the end of its life. 
The density in the core is so high that the electrons there are relativistically 
degenerate. Capture of these electrons by Ne and Mg nuclei reduces the 
electron degeneracy pressure and triggers the collapse of the core. 
In contrast, a star of $>9 M_\odot$ develops an Fe core, which collapses 
when thermal energy is lost due to photo-dissociation of Fe-group nuclei. 
In both cases, the inner core bounces due to the repulsive nuclear force at 
very short range when supra-nuclear density is reached. 
This bounce launches a shock wave 
into the still-collapsing outer core. However, the shock quickly loses energy 
on its way out by dissociating nuclei into free nucleons and is stalled before 
exiting the outer core. The inner core is now a proto-NS and material falling 
onto it releases the gravitational binding energy by emitting mostly $\nu_e$ 
and $\bar\nu_e$. Some of these $\nu_e$ and $\bar\nu_e$ are captured by 
neutrons and protons through the forward reactions in Eqs.~(\ref{eq-np}) 
and (\ref{eq-pn}), respectively, to heat the material behind the stalled shock.
In some cases, this neutrino heating provides sufficient energy to
revive the shock, which proceeds to make an explosion. 
This is the so-called neutrino-driven SN mechanism \cite{bethe}.

The above SN mechanism has been consistently demonstrated by several 
groups for a star of $8.8M_\odot$ \cite{mayle,kitaura,fischer}. 
However, the same mechanism is 
harder to operate in more massive stars, which have more extended 
envelopes with larger gravitational binding energies. 
For these more massive stars, the shock 
is required to do extra work and sometimes neutrinos fail to deliver an explosion. 
At the present time, whether neutrino-driven explosion works for stars of 
$> 9 M_\odot$ and if so, how exactly it works are under intense study by 
many SN modelers around the world \cite{janka}. An exploratory study by the Garching 
group in Germany found that whether neutrino-driven explosion works is not 
a simple function of the SN progenitor mass \cite{janka}. In addition, 
when a neutrino-driven SN occurs, it takes $\sim 0.1$ to $\sim 1$~s and
the explosion energy varies from $\sim 10^{50}$ to $\sim 10^{51}$~ergs.
Neither the time nor the energy of explosion is a monotonic function 
of the progenitor mass.

A successful explosion typically leaves behind an NS of a few $M_\odot$, 
while a failed SN produces a BH that swallows the entire 
progenitor star most of the time. In some cases, an accretion
disk may form around the BH and powers a jet that drives an explosion
ejecting part of the progenitor star. In rare cases, this jet-driven mechanism
gives rise to the so-called hypernovae associated with long gamma-ray bursts
\cite{mac}.

\subsection{SN neutrino emission}
The $\nu_e$ and $\bar\nu_e$ potentially driving the explosion are emitted 
dominantly through the reverse reactions in Eqs.~(\ref{eq-np}) and (\ref{eq-pn}), 
respectively, by material falling onto the proto-NS. The so-called accretion phase 
associated with this emission
lasts $\sim 0.1$ to $\sim 1$~s. The gravitational binding energy of the proto-NS 
itself [see Eq.~(\ref{eq-eg})] is released in $\nu_e$, $\bar\nu_e$, $\nu_\mu$, $\bar\nu_\mu$, 
$\nu_\tau$, and $\bar\nu_\tau$ during the so-called cooling phase, for which 
the important neutrino production mechanisms are processes such as
$e^++e^-\to\nu+\bar\nu$. The cooling phase lasts
$\sim 10$~s because neutrinos must diffuse out of the extremely hot and dense 
interior of the proto-NS. Detection of the neutrino burst from SN 1987A
\cite{hirata,bionta}, which lasted $\approx 13$~s, confirmed this overall picture of 
SN neutrino emission. 

\begin{figure}[H]
\hskip-0.5cm
\includegraphics[scale=0.4]{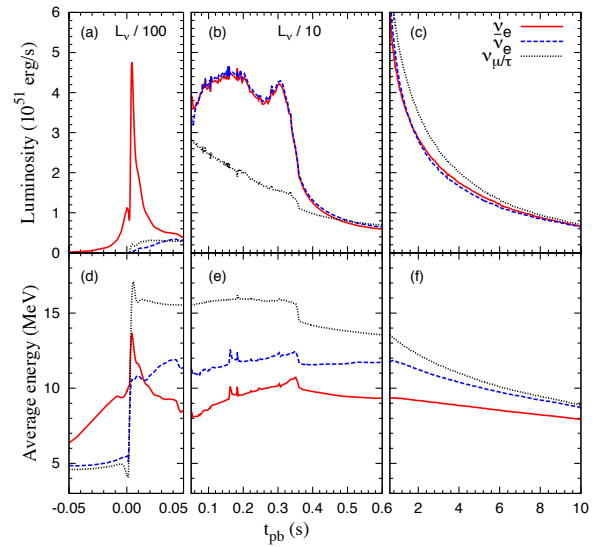}
\caption{Evolution of neutrino luminosities and average neutrino energies as functions of time for
an $18 M_\odot$ SN model \cite{fischer,wu}. Panels (a) and (d) correspond to the shock-breakout phase, 
panels (b) and (e) the accretion phase, and panels (c) and (f) the cooling phase. 
(Figure from Ref.~\cite{wu})
} %图题
\label{fig-nu}
\end{figure}

The characteristics of neutrino emission differ greatly between the 
accretion and cooling phases. 
Figure~\ref{fig-nu} shows the evolution of neutrino luminosities and 
average neutrino energies as functions of time for an $18M_\odot$ SN model
\cite{fischer,wu}.
During the accretion phase (Fig.~\ref{fig-nu}b,
time post core bounce $t_{\rm pb}\sim 0.05$--0.6~s), 
the $\bar\nu_e$ luminosity $L_{\bar\nu_e}$ is approximately the same as the $\nu_e$ 
luminosity $L_{\nu_e}$. Both follow nearly the same time evolution and are 
much higher than the luminosity $L_{\nu_x} \approx L_{\bar\nu_x}$ 
$(x = \mu, \tau)$ of any other species. In addition, the average neutrino
energies (Fig.~\ref{fig-nu}e) follow a clear hierarchy
$\langle E_{\nu_e}\rangle<\langle E_{\bar\nu_e}\rangle<
\langle E_{\nu_x}\rangle\approx\langle E_{\bar\nu_x}\rangle$, with
$\langle E_{\nu_e}\rangle\approx 8$--11~MeV, 
$\langle E_{\bar\nu_e}\rangle\approx 11$--12~MeV, and
$\langle E_{\nu_x}\rangle\approx\langle E_{\bar\nu_x}\rangle\approx 16$--14~MeV.
In contrast, $L_{\nu_x}\approx L_{\bar\nu_x}$ is close to 
$L_{\nu_e} \approx L_{\bar\nu_e}$ during the cooling phase 
(Fig.~\ref{fig-nu}c, $t_{\rm pb} > 0.6$~s),
with all species having nearly the same luminosity eventually.
The average neutrino energies monotonically decrease with time 
(Fig.~\ref{fig-nu}f),
and also become less distinctive from each other. While
$\langle E_{\nu_e}\rangle$ remains the lowest, the difference between
$\langle E_{\bar\nu_e}\rangle$ and
$\langle E_{\nu_x}\rangle\approx\langle E_{\bar\nu_x}\rangle$ becomes smaller
and smaller. Note that Figs.~\ref{fig-nu}a and 
\ref{fig-nu}d correspond to the so-called shock-breakout phase.
As the shock breaks through the neutrino-trapping surface formed by nuclei of the Fe 
group, the protons released from the dissociation of these nuclei rapidly capture electrons 
to produce a strong $\nu_e$ pulse. Therefore, the shock-breakout phase is characterized 
by powerful emission of predominantly $\nu_e$. 

For SNe with different progenitors, the shock-breakout pulse is a common feature that
signifies the launch of the prompt shock. A $\nu_e$ pulse with
$L_{\nu_e}\approx (4$--$5)\times10^{53}$~erg/s and a width of $\sim 10$~ms
is typical of all SNe. For those SNe with neutrino-driven explosion, the duration of the
accretion phase depends on the progenitor. For example, this phase lasts $\sim 0.6$~s
for the $18M_\odot$ model described above, which is $\sim 3$ times longer than that
for the $8.8M_\odot$ model discussed in Ref.~\cite{janka}. The cooling phase is similar for
all SNe that leave behind NS remnants. For SNe producing BHs, when the BH forms
depends on the progenitor structure and the nuclear equation of state. If a BH forms 
during the first $\sim 10$~s after core bounce, neutrino emission from the proto-NS is
abruptly terminated (e.g., \cite{fis}). However, if there is an accretion disk 
surrounding the BH, significant neutrino emission from this disk
might continue for some time (e.g., \cite{pop}).

\subsection{Neutrino-driven winds and heavy-element synthesis}
\label{sec-wind}
Subsequent to a successful SN explosion, material in the vicinity of the proto-NS is still 
heated by $\nu_e$ and $\bar\nu_e$ through the forward reactions in Eqs.~(\ref{eq-np}) 
and (\ref{eq-pn}), 
respectively. When this material acquires sufficient energy from neutrino heating, it 
overcomes the gravitational potential of the proto-NS and escapes as the so-called 
neutrino-driven wind. The neutron-to-proton ratio in the wind is determined by the 
competition between neutron production by $\bar\nu_e$ and proton production by 
$\nu_e$ through the same reactions that provide the heating \cite{qian1993,qian1996}. 
The rates for these reactions at radius $r$ are
\begin{align}
\lambda_{\bar\nu_ep}=\frac{L_{\bar\nu_e}}{4\pi r^2}
\frac{\langle\sigma_{\bar\nu_ep}\rangle}{\langle E_{\bar\nu_e}\rangle}
\propto L_{\bar\nu_e}\left(\frac{\langle E_{\bar\nu_e}^2\rangle}
{\langle E_{\bar\nu_e}\rangle}-2\Delta\right),\\
\lambda_{\nu_en}=\frac{L_{\nu_e}}{4\pi r^2}
\frac{\langle\sigma_{\nu_en}\rangle}{\langle E_{\nu_e}\rangle}
\propto L_{\nu_e}\left(\frac{\langle E_{\nu_e}^2\rangle}
{\langle E_{\nu_e}\rangle}+2\Delta\right),
\end{align}
where the angular brackets indicate averaging over the relevant neutrino energy 
spectrum, $\sigma_{\bar\nu_ep}\propto(E_{\bar\nu_e}-\Delta)^2$ and 
$\sigma_{\nu_en}\propto(E_{\nu_e}+\Delta)^2$ are the cross sections for the 
corresponding reactions, and we have ignored terms proportional to $\Delta^2$. 
The neutron-to-proton ratio in the wind can be estimated as
\begin{equation}
n/p\approx\frac{\lambda_{\bar\nu_ep}}{\lambda_{\nu_en}}\approx
\frac{L_{\bar\nu_e}}{L_{\nu_e}}
\left(\frac{\epsilon_{\bar\nu_e}-2\Delta}{\epsilon_{\nu_e}+2\Delta}\right),
\end{equation}
where $\epsilon_\nu\equiv\langle E_\nu^2\rangle/\langle E_\nu\rangle$.
For $L_{\bar\nu_e}\approx L_{\nu_e}$, $n/p>1$ requires
$\epsilon_{\bar\nu_e}-\epsilon_{\nu_e}>4\Delta$.

Because neutrino energy spectra are determined by neutrino opacities in the 
surface layers of the proto-NS, whether the wind is neutron rich is sensitive to 
neutrino interactions in hot and dense matter. Using various approximate 
neutrino opacities, earlier studies found the wind to be mostly neutron rich 
(e.g., \cite{qian1996,woosley1994}), whereas later ones obtained only proton-rich winds 
(e.g., \cite{hudepohl}) with neutrino emission parameters similar to those shown in 
Fig.~\ref{fig-nu}. Recently, two groups \cite{martinez,roberts2012} 
studied $\nu_e$ and $\bar\nu_e$ opacities in some detail and found 
that the wind may be neutron rich for a significant period of time.

A number of groups (e.g., 
\cite{qian1996,woosley1994,witti,takahashi,hoffman,wanajo2001,thompson,roberts2010}) 
have studied other conditions such
as the entropy and expansion timescale in the neutrino-driven wind and 
surveyed the resulting nucleosynthesis. The general consensus is that 
elements such as Sr, Y, and Zr with $A\sim 90$ can be readily produced 
for somewhat neutron-rich winds. It is likely that the production can be 
extended to Pd and Ag with $A\sim 110$. However, these nuclei are made 
mainly through a quasi-equilibrium process \cite{woosley1992} involving 
$(n,\gamma)$, $(p,\gamma)$, $(n,p)$, $(\alpha,\gamma)$, $(\alpha,n)$, 
$(\alpha,p)$, and their reverse reactions, in contrast to the classical $r$ 
process where neutron capture plays a dominant role. With extreme 
conditions such as associated with a massive proto-NS \cite{wanajo2013}, a classical 
$r$ process can occur to produce nuclei up to $A\sim 130$. However, 
it is very difficult to justify conditions for making $r$-process nuclei with 
$A\sim 195$ in the wind.

\subsection{Neutrino-induced $r$ process in helium shells}
In addition to driving a neutron-rich wind, neutrino interactions in SNe 
can provide neutrons in other ways as well. For example, neutral-current 
reactions on $^4$He nuclei can produce neutrons through 
$^4{\rm He}(\nu,\nu n)^3{\rm He}(n, p)^3{\rm H}$ and 
$^4{\rm He}(\nu,\nu p)^3{\rm H}$ followed by $^3{\rm H}(^3{\rm H},2n)^4{\rm He}$. 
In the helium (He) shell of an early SN, these neutrons are captured by the few 
$^{56}$Fe nuclei present in the birth material of the progenitor, but not by the 
predominant $^4$He nuclei. This scenario was proposed as a model for the 
$r$ process \cite{epstein}. It was critically examined by Ref.~\cite{woosley1990}, 
which constrained it to be viable 
only for some special metal-poor SNe with He shells at very small radii and hence, 
exposed to large neutrino fluxes for neutron production.

A recent study \cite{banerjee2011} 
reexamined the above scenario for a neutrino-induced $r$ process. 
Using updated models of metal-poor massive stars, it found that the $^3$H nuclei 
produced by neutral-current neutrino reactions on $^4$He lead to production of 
$^7$Li through $^4{\rm He}(^3{\rm H},\gamma)^7{\rm Li}$ instead of generating 
neutrons as in the original scenario. However, the charged-current reaction
\begin{equation}
\bar\nu_e + {^4{\rm He}}\to{^3{\rm H}} + n + e^+
\label{eq-nua}
\end{equation}
may provide a new neutron source, especially in the presence of 
$\bar\nu_e\rightleftharpoons\bar\nu_x$ oscillations. The reaction in Eq.~(\ref{eq-nua}) 
has a threshold of 21.6 MeV, which is significantly above the average $\bar\nu_e$ 
energy in the absence of flavor oscillations. Earlier SN neutrino 
transport calculations (e.g., \cite{woosley1994}) gave a very hard $\bar\nu_x$ spectrum 
with $\langle E_{\bar\nu_x}\rangle\sim 20$--25~MeV. More recent calculations also 
showed that the emission spectrum of $\bar\nu_x$ is significantly harder than 
that of $\bar\nu_e$ at least for a few seconds (see Fig.~\ref{fig-nu}). 
For an inverted neutrino mass hierarchy, 
$\bar\nu_e\rightleftharpoons\bar\nu_x$ oscillations can occur before neutrinos 
reach the He shell, thereby giving rise to a harder effective $\bar\nu_e$ spectrum 
for neutron production.

The neutron production rate per $^4$He nucleus is
\begin{equation}
\lambda_{\bar\nu_e\alpha,n} =\frac{1}{4\pi r^2}\left[
\frac{L_{\bar\nu_e}\langle\sigma_{\bar\nu_e\alpha,n}\rangle}{\langle E_{\bar\nu_e}\rangle}
\right]_{\rm eff}\propto\frac{(L_{\bar\nu_e}T_{\bar\nu_e}^p)_{\rm eff}}{r^2},
\end{equation}
where $\langle\sigma_{\bar\nu_e\alpha,n}\rangle$ is the cross section for the 
charged-current reaction in Eq.~(\ref{eq-nua}) averaged over the $\bar\nu_e$ spectrum, 
the subscript ``eff'' denotes effective quantities for $\bar\nu_e$ in the presence of
$\bar\nu_e\rightleftharpoons\bar\nu_x$ oscillations, $T_{\bar\nu_e}$ is the 
temperature for a Fermi-Dirac spectrum with zero chemical potential, and the power index 
$p$ is $\sim 5$--6. Using the neutrino emission spectra of Ref.~\cite{woosley1994} and invoking 
$\bar\nu_e\rightleftharpoons\bar\nu_x$ oscillations, Ref.~\cite{banerjee2011} showed that an 
$r$ process can occur and produce nuclei up to $A > 200$ in the He shell of an $11 M_\odot$ 
SN model with an initial metallicity of 
${\rm [Fe/H]}\equiv\log{\rm (Fe/H)}-\log{\rm (Fe/H)}_\odot\sim -4.5$. 

\begin{figure}[H]
\centering
\includegraphics[scale=0.25]{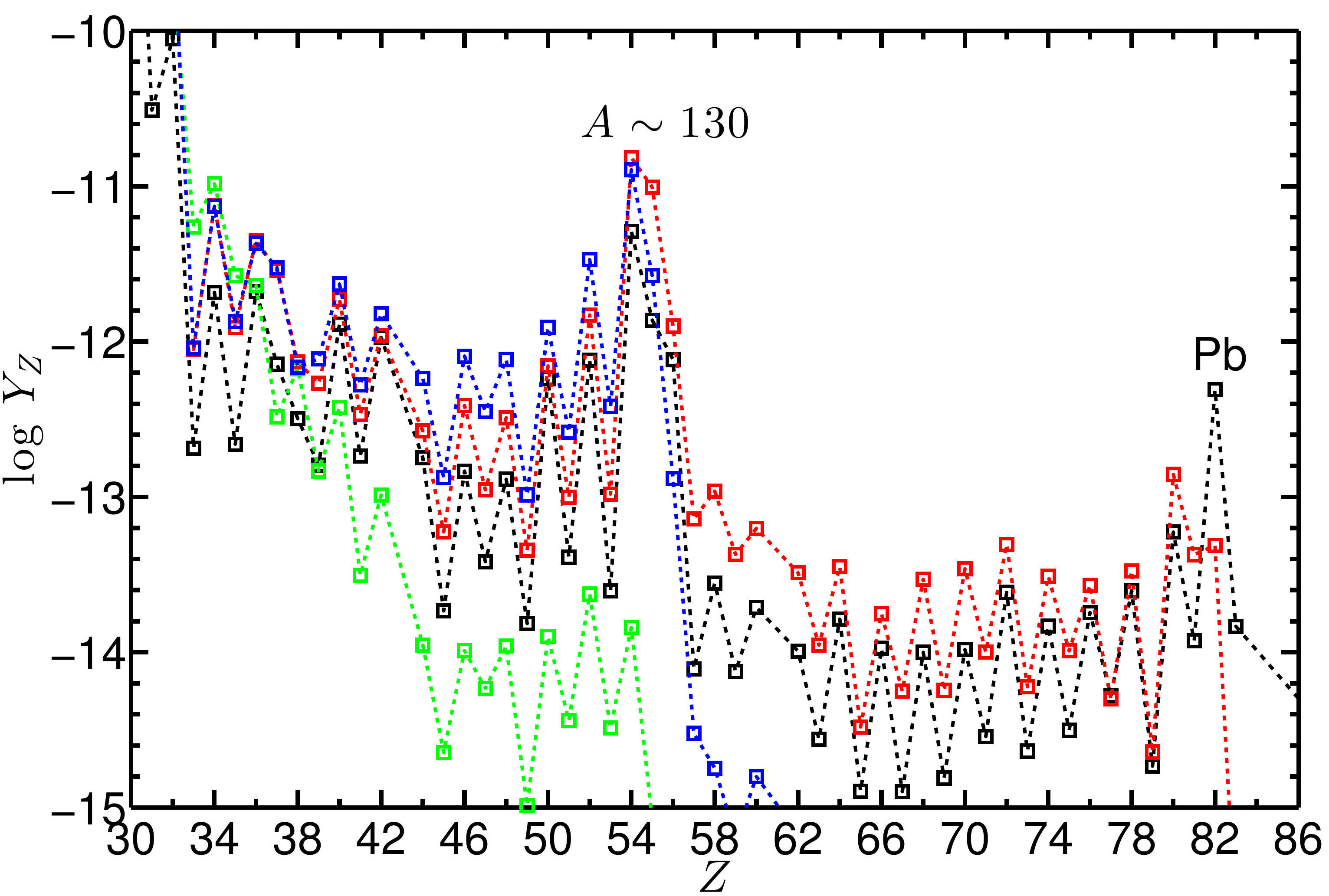}
\caption{Effect of SN progenitor metallicity on neutrino-induced 
$r$-process nucleosynthesis in He shells \cite{banerjee2011,banerjee2016}.
Final elemental abundance patterns are shown as functions of atomic number $Z$
for $11 M_\odot$ models with metallicities of ${\rm [Fe/H]}\sim -5$ (black), $-4$ 
(red and blue), and $-3$ (green). The two models with [Fe/H]~$\sim -4$ have different
abundances of $^{28}$S and $^{32}$Si in the He shell ($\sim 10$ times
smaller for the red curve). (Figure from Ref.~\cite{banerjee2016})
} %图题
\label{fig-he}
\end{figure}

The neutron density obtained in the He shell is approximately determined by
the competition between neutrino-induced production and capture by nuclei such as $^{56}$Fe
in the birth material. As the metallicity of the SN progenitor increases, more $^{56}$Fe nuclei
are available to capture neutrons, which results in lower neutron densities and less efficient 
production of nuclei heavier than $^{56}$Fe. Consequently,
the above neutrino-induced $r$ process ceases to operate at ${\rm [Fe/H]}\gtrsim -3$. 
This dependence on metallicity \cite{banerjee2016} is illustrated in Fig.~\ref{fig-he}.

\subsection{NS mergers and the $r$ process}
\label{sec-nsm}
The NS left behind by an SN is a great source of neutron-rich material. This source
can be tapped to drive a robust $r$ process during mergers of an NS with another 
NS or a BH. The progenitor system of such a merger is a binary consisting of two 
massive stars, which explode as SNe without disrupting the system. Energy loss 
through radiation of gravitational waves leads to the eventual merger of the two 
compact remnants left by the SNe. Pioneering work on $r$-process nucleosynthesis 
during decompression of cold NS matter was carried out in Ref.~\cite{lattimer}. 
More recently, detailed hydrodynamic simulations of an NS-NS merger were performed 
(e.g., \cite{korobkin,bauswein}). It was shown that $r$-process nuclei with $A\gtrsim 130$ 
including thorium and uranium are produced in the extremely neutron-rich ejecta. 
In addition, the less neutron-rich material ejected from the accretion disk surrounding
the merger remnant has similar nucleosynthesis (e.g., \cite{just}) to the neutron-rich 
neutrino-driven winds from a proto-NS (see Sec.~\ref{sec-wind}).

The production of $r$-process nuclei with $A\gtrsim 130$ and the associated
production of lighter nuclei in an NS-NS merger received strong support from
the observations of such an event, GW170817, through gravitational waves \cite{abbott} 
and electromagnetic radiation. In this regard, the most dramatic observation
of this event is the detection of the so-called kilonova \cite{smartt}, which was 
powered by the decay of the nuclei synthesized by the $r$ process (e.g., \cite{kasen}). 

\section{Conclusions}
Eight decades after Baade and Zwicky proposed the connection between SNe 
and NS formation, we are still figuring out the mechanisms through which SNe occur. 
It has been shown that neutrino-driven explosion works for stars of $\sim 8$--$9 M_\odot$. 
While neutrinos may also play important roles in explosions of more massive stars, the 
detailed mechanisms in these cases are rather uncertain but under intense investigation 
at the present time. Nevertheless, it appears that SNe from stars of $\sim 8$--$100 M_\odot$ 
have a wide range of neutrino signals, explosion energies, nucleosynthesis products, and 
compact remnant (NS or BH) masses.

With formation of an NS and the associated profuse emission of neutrinos, SNe
can provide neutrons for making heavy nuclei, especially through the $r$ process, 
in several ways. First of all, neutrino-driven winds from a proto-NS can be neutron 
rich due to the dominance of the forward reaction in Eq.~(\ref{eq-pn}). These winds can 
produce elements from Sr, Y, Zr ($A\sim 90$) up to Pd and Ag ($A\sim 110$) through a 
quasi-equilibrium process, and for the most favorable conditions, make nuclei up to 
$A\sim 130$ through the $r$ process. In addition, $\bar\nu_e$ can produce neutrons 
through the reaction in Eq.~(\ref{eq-nua}). This may give rise to a neutrino-induced $r$ process 
in the He shell of an early SN where neutrons are captured by the few $^{56}$Fe nuclei 
present in the birth material of the progenitor star. As neutron production is sensitive to 
the effective $\bar\nu_e$ energy spectrum in the He shell, this neutrino-induced $r$ 
process can be enhanced greatly by $\bar\nu_e\rightleftharpoons\bar\nu_x$ oscillations, 
the occurrence of which most likely requires an inverted neutrino mass hierarchy. In the 
optimal case, nuclei with $A > 200$ can be produced. However, the neutrino-induced 
$r$ process ceases to operate when metallicities of SN progenitors exceed 
${\rm [Fe/H]}\sim -3$. Finally, a small fraction of binary systems consisting of two 
massive stars can evolve into NS-NS or NS-BH binaries after surviving two SNe. 
Cold NS matter ejected from mergers of the two compact remnants in such binaries 
serves as the best site for making $r$-process nuclei with $A\gtrsim 130$ including 
thorium and uranium.

\vspace*{2mm} \Acknowledgements{\bahao
I thank Projjwal Banerjee, Tobias Fischer, Wick Haxton, Alexander Heger, 
Gabriel Mart{$\acute\imath$}nez-Pinedo, and Meng-Ru Wu for fruitful collaboration. 
This work was supported in part by US DOE Grant No. DE-FG02-87ER40328.}

%%%%%%%%%%%%%%%%%%%%%%%%%%%%%%%%%%%%%%%%%%%%%%%%%%%%%%%%%%%%
%% Text of article.
%%%%%%%%%%%%%%%%%%%%%%%%%%%%%%%%%%%%%%%%%%%%%%%%%%%%%%%%%%%%
%    Section headings
\renewcommand{\baselinestretch}{1.08} \baselineskip 12.2pt\parindent=10.8pt

\renewcommand{\thefootnote}

\end{multicols}

\end{document}